\documentclass[conference]{IEEEtran}
\IEEEoverridecommandlockouts

\usepackage{cite}
\usepackage{amsmath,amssymb,amsfonts}
\usepackage{graphicx}
\usepackage{textcomp}
\usepackage[dvipsnames]{xcolor}
\usepackage[final]{microtype}
\usepackage[all]{nowidow}
\usepackage[keeplastbox]{flushend}

\usepackage{booktabs}
\usepackage{hyperref}

\begin{document}

\title{A Controlled Study of Memory Hierarchy Transitions in Quantum
Circuit Simulation on Apple M4 Pro Unified Memory Architecture}

\author{\IEEEauthorblockN{Gyan Pratipat}
\IEEEauthorblockA{\textit{Ira A. Fulton Schools of Engineering} \\
\textit{Arizona State University} \\
Tempe, AZ, USA \\
\texttt{gpratipa@asu.edu}}}

\maketitle

\begin{abstract}
State-vector quantum circuit simulation is memory-bandwidth bound, but
the relative contributions of memory hierarchy, access pattern, and
hardware parallelism are difficult to separate on conventional
hardware.  We study this on the Apple
M4~Pro unified memory architecture, where CPU and GPU share identical
physical DRAM (${\sim}224$~GB/s for both), eliminating
memory-technology and interconnect confounds.  Using a thermally
isolated methodology across 11 backends on GHZ and QFT circuits from 3
to 30 qubits, we identify a 4.3--4.5$\times$ timing discontinuity at
the 28$\rightarrow$29 qubit transition.  Two controlled experiments
localize its cause: doubling amplitude precision moves the
discontinuity exactly one qubit earlier, recurring at the identical
4.29~GB working set, and a cross-simulator comparison against Qiskit
Aer and Qulacs shows that only kernels materializing a full transposed
buffer are susceptible.  Byte footprint determines where a susceptible
kernel breaks down; access pattern determines whether it is susceptible
at all.  Measured GPU speedups of 3--10$\times$ exceed a STREAM-style
bandwidth prediction of 1.85$\times$.  They are within-framework
comparisons to isolate the device variable and shall not be treated as
universal GPU-over-CPU speed-up.
\end{abstract}


\section{Introduction}

State-vector quantum circuit simulation is a dominant method for
validating quantum algorithms on classical hardware.  The exponential
memory requirement is well understood: an $n$-qubit system requires
a state vector of $2^n$ complex amplitudes, consuming $2^n \times 8$
bytes in single precision requiring 8.59~GB at 30 qubits and exceeding the DRAM capacity of consumer and laptop-class hardware at 40 qubits.  As the qubit count grows, simulation becomes increasingly DRAM-bound as the state vector exceeds on-chip cache capacity.

While the role of the memory hierarchy is well understood in general, its impact on large-scale state-vector quantum simulation remains less clearly characterized. In particular, it is unclear at what problem sizes memory hierarchy effects manifest as observable performance discontinuities, how severe these transitions are in practice, and to what extent they depend on the access patterns of the simulation algorithm. These questions have direct implications for hardware selection and the design of representative benchmarks.

\subsection{The Attribution Problem}

Prior work establishes qualitatively that state-vector simulation is
memory-bandwidth bound~\cite{9910084,11333848}; SV-Sim~\cite{9910084}
places arithmetic intensity below 0.5~FLOP/byte, and
Queen~\cite{queen2024} and DiaQ~\cite{diaq2024} confirm this on NVIDIA
hardware via Roofline analysis.  Attribution, however, remains
unresolved.  QARN~\cite{11333848} reports 18$\times$ GPU speedup but
measures only CPU STREAM bandwidth (12--15~GB/s on i7, 65--75~GB/s on
Xeon), inferring the GPU figure from datasheets; the resulting
${\sim}$12--13$\times$ bandwidth ratio cannot account for an
18$\times$ speedup, leaving the operative variable unidentified.

Conventional CPU-to-GPU comparisons simultaneously vary four
architectural factors: memory bandwidth, memory technology (DDR5 vs.\
GDDR6), controller architecture, and compute parallelism.  When all
four vary simultaneously, observed speedup cannot be cleanly attributed
to any single factor.  The Apple M4~Pro UMA addresses this confound:
CPU and GPU share one physical LPDDR5X DRAM pool, measured here at
${\sim}$224~GB/s STREAM-style bandwidth for both (Table~\ref{tab:bandwidth}),
leaving memory-access parallelism and access-pattern structure as the
differentiating variables.

\subsection{Contributions}

\begin{enumerate}
    \item \textbf{Roofline Analysis:} We derive arithmetic intensity (AI)
    for all common quantum gate implementations and show all backends have
    AI $\leq 0.38$~FLOP/byte, well below the ridge point for any plausible peak compute on this hardware,
    confirming structural memory-boundedness.

    \item \textbf{DRAM Bandwidth Cliff Characterization:} We identify
    and characterize a reproducible 4.3--4.5$\times$ timing discontinuity
    at the 28$\rightarrow$29 qubit transition, thermally isolated and
    cross-validated across GHZ and QFT circuit classes.  Direct-index
    backends maintain ${\sim}$2$\times$ scaling at all qubit counts,
    consistent with scale-invariant DRAM-limited behavior throughout.

    \item \textbf{STREAM-Style Bandwidth vs.\ Simulation Speedup:} Despite
    a STREAM-style measurement predicting only 1.85$\times$ GPU speedup (MLX CPU
    119.9~GB/s vs.\ MLX GPU 221.9~GB/s), all three algorithm classes
    exceed this prediction: tensordot achieves 3.1--4.1$\times$,
    flat-index 3.5--5.9$\times$, and direct-index 6--10$\times$.  This demonstrates that
    peak streaming bandwidth is an insufficient predictor for
    non-contiguous memory access patterns, with the gap widening as
    access irregularity increases.
\end{enumerate}

\subsection{Paper Organization}

Section~\ref{sec:background} presents background and related work.
Section~\ref{sec:methodology} describes experimental methodology.
Section~\ref{sec:results} reports results: Roofline position
(Section~\ref{sec:roofline_position}), the DRAM cliff
(Section~\ref{sec:cliff}), and the STREAM-style bandwidth versus simulation speedup
analysis (Section~\ref{sec:bandwidth_speedup}).
Section~\ref{sec:discussion} discusses implications and limitations.
Section~\ref{sec:conclusion} concludes.

\section{Background and Related Work}
\label{sec:background}

\subsection{State-Vector Simulation}

In gate-based quantum computation, an $n$-qubit system is represented
by a state vector of $2^n$ complex amplitudes
$\{\nu_0, \nu_1, \ldots, \nu_{2^n-1}\}$, each a complex64 value
(8 bytes, single precision).  Applying a single-qubit gate $G$ to
qubit $t$ transforms pairs of amplitudes differing only in bit $t$,
touching exactly half the state vector per gate
($2^{n-1}$ amplitude pairs)~\cite{viamontes2003gate,viamontes2009quantum}.

The naive implementation materializes the full $2^n \times 2^n$ gate
matrix via Kronecker products, requiring $\mathcal{O}(4^n)$ memory.
Optimised implementations avoid this by operating directly on state
vector element pairs via direct index manipulation~\cite{qhipster2016,
yu2025qvecopt,11333848}, or through tensordot-based approaches,
achieving $\mathcal{O}(2^n)$ working memory.

State vectors admit two equivalent representations.  The
\emph{flat representation} stores all $2^n$ amplitudes as a
single contiguous array of shape $[2^n]$.  The \emph{tensor
representation} reshapes the state vector into a rank-$n$ tensor
of shape $[2, 2, \ldots, 2]$, where each axis corresponds to one qubit;
applying a gate to qubit $t$ reduces to a tensor contraction along
axis $t$.  The tensor representation enables clean gate application
code but generates less predictable memory access patterns than direct
index manipulation.  This distinction has performance consequences
at the DRAM bandwidth cliff, characterized in
Section~\ref{sec:cliff}.

\subsection{The Roofline Model}

The Roofline model~\cite{williams2009roofline} gives the upper-bound
achievable performance as:
\begin{equation}
    P = \min\!\left(P_{\text{peak}},\
        \text{AI} \times B_{\text{peak}}\right)
\end{equation}
where $P$ is performance (GFLOP/s), $P_{\text{peak}}$ is peak compute
throughput, $B_{\text{peak}}$ is peak memory bandwidth (GB/s), and
Arithmetic Intensity (AI) is floating-point operations per byte of
memory traffic (FLOP/byte).  The ridge point
$\text{AI}^* = P_{\text{peak}} / B_{\text{peak}}$ separates the
memory-bound ($\text{AI} < \text{AI}^*$) from the compute-bound
regime.

SV-Sim~\cite{9910084} establishes AI $< 0.5$~FLOP/byte for quantum
simulation.  Queen~\cite{queen2024} provides Roofline analysis for
30-qubit operations on NVIDIA GPUs, confirming deep memory-boundedness.
DiaQ~\cite{diaq2024} characterizes strided memory access in
state-vector simulation and addresses it by transforming strided into
linear accesses.  PIMutation~\cite{pimutation2025} applies Roofline
analysis to motivate PIM acceleration for quantum simulation, further
establishing memory-boundedness across 16--32 qubits.  Faj
et al.~\cite{faj2023} characterize GPU acceleration for state-vector
simulation, identifying memory bandwidth as the primary bottleneck.
Cache-to-DRAM performance transitions on conventional NUMA systems have been
characterized by Molka et al.~\cite{molka2009memory}; this paper extends the
Roofline approach to unified-memory hardware and quantifies the equivalent
throughput discontinuity on Apple Silicon.

\subsection{Existing Quantum Simulators}

A systematic benchmarking study of 24 state-vector simulators on HPC
hardware is provided by Jamadagni et al.~\cite{jamadagni2024}, covering
single-thread, multithread, and GPU configurations.  Primary simulators
include qHiPSTER~\cite{qhipster2016}, QuEST~\cite{quest2019},
Qulacs~\cite{qulacs2021}, Qiskit Aer~\cite{qiskit2024},
cuQuantum~\cite{cuquantum2023}, SV-Sim~\cite{9910084}, and
QARN~\cite{11333848}.  Kumaresan et al.~\cite{kumaresan2026} benchmark
GPU-accelerated simulation with empirical backend selection, gate
fusion, and adaptive precision.  Vallero et al.~\cite{vallero2025}
evaluate GPU performance of state-vector and tensor-network simulation
on NVIDIA hardware, identifying cache saturation effects at 13--22
qubits.  osxQuantum~\cite{osxquantum2026} is a Metal-accelerated
state-vector simulator for Apple Silicon built on the MLX framework.
We systematically characterize the memory hierarchy behavior,
the access-pattern-dependent throughput discontinuity, and the
STREAM-style bandwidth versus simulation speedup gap of state-vector simulation on Apple
Silicon unified memory.

\subsection{Unified Memory Architecture}

In conventional GPU systems, CPU and GPU maintain separate memory
pools---host DRAM (${\sim}$75~GB/s for a typical Xeon) versus GPU VRAM
(${\sim}$864~GB/s for NVIDIA L40)~\cite{9910084} and PCIe
interconnect (${\sim}$32~GB/s bidirectional).  These pools differ in
technology (DDR5 vs.\ GDDR6), controller, and physical location.

Apple Silicon UMA eliminates these confounds.  CPU and GPU share one
physical LPDDR5X DRAM pool fabricated on-chip~\cite{apple2024m4pro,
apple2024m4pro_specs}.  Our measurements confirm both backends achieve
${\sim}$224~GB/s STREAM-style bandwidth (Table~\ref{tab:bandwidth}).  The
M4~Pro GPU comprises 20 shader cores dispatching thousands of
concurrent Metal threads~\cite{apple2024m4pro_specs}, versus the
14-core CPU running at most 14 concurrent execution contexts.  This
hardware parallelism difference governs effective memory throughput for
non-contiguous access patterns, independently of peak DRAM bandwidth.

\section{Experimental Methodology}
\label{sec:methodology}

\subsection{Hardware Configuration}

All experiments were conducted on an Apple MacBook Pro (Mac16,7) with
an Apple M4~Pro SoC: 14-core CPU (10 performance + 4 efficiency), 20-core
GPU, and 48~GB LPDDR5X unified memory~\cite{apple2024m4pro_specs,
apple2024m4pro}.  Theoretical peak bandwidth is 273~GB/s.
H\"{u}bner et al.~\cite{hubner2025} provide a systematic HPC
characterization of M1--M4 SoCs; our work complements their analysis
by targeting quantum simulation workloads on the M4~Pro specifically.

Memory bandwidth was characterized using a STREAM-style probe~\cite{mcCalpin1995}: 512~MB
float32 arrays, 10 kernel-warmup passes to allow JIT compilation, then
5 independently timed read+write passes.  For MLX GPU, trial~1 of the
timed passes was excluded due to GPU frequency ramp-up; the reported
GPU value is the mean of trials~2--5.  Results are in
Table~\ref{tab:bandwidth}.

\begin{table}[h]
\caption{Measured sustained STREAM-style bandwidths on M4~Pro (Mac16,7).
512~MB float32, 5 timed passes after 10 kernel-warmup passes.
$^\dagger$GPU trial~1 excluded (frequency ramp-up); $N=4$.}
\label{tab:bandwidth}
\small
\begin{tabular}{lrrr}
\toprule
\textbf{Backend} & \textbf{BW (GB/s)} &
\textbf{$\sigma$} & \textbf{\% peak} \\
\midrule
JAX CPU   & 224.7 & $\pm$0.1   & 82.3\% \\
MLX GPU   & 221.9 & $\pm$1.4 & 81.3\% \\
MLX CPU   & 119.9 & $\pm$0.1   & 43.9\% \\
\bottomrule
\end{tabular}
\end{table}

JAX CPU and MLX GPU achieve near-identical STREAM-style bandwidth,
confirming shared physical DRAM.  MLX CPU reaches only 44\% of peak,
reflecting overhead in MLX's CPU dispatch path relative to JAX's
XLA-compiled kernels.

MLX~\cite{mlx2023} is Apple's open-source array framework for Apple
Silicon; GPU execution is backed by Metal~\cite{apple_metal}.
JAX~\cite{jax2018} routes CPU execution through XLA compilation
and Apple's AMX units.  All backends access identical physical DRAM.
An experimental \texttt{jax-metal} plugin provides Metal GPU
support for JAX~\cite{apple_jax_metal}, but it explicitly does
not support \texttt{complex64} or \texttt{complex128} datatypes
and is therefore incompatible with state-vector simulation;
it is excluded from this study.
\subsection{Benchmark Circuits}

\begin{enumerate}
\item \textbf{Greenberger-Horne-Zeilinger (GHZ):} One Hadamard gate on qubit~0 followed by $n-1$
CNOT gates, producing $(|00\ldots0\rangle + |11\ldots1\rangle)/\sqrt{2}$.
Maximal entanglement prevents tensor-network shortcuts; every gate
touches half the state vector~\cite{11333848,diaq2024,jamadagni2024}.

\item \textbf{Quantum Fourier Transform (QFT):} $\mathcal{O}(n^2)$ gates (Hadamard, controlled-phase,
SWAP), 480 gates at 30 qubits.  Used to cross-validate cliff location
under a qualitatively different gate structure with $16\times$ higher
gate count.
\end{enumerate}

Correctness was verified against Qiskit Aer~\cite{qiskit2024} statevector
at small qubit counts: GHZ across 5 qubit counts (maximum amplitude
deviation $< 10^{-6}$), QFT across 5 backends at 3--4 qubits (all PASS).

\subsection{Simulation Backends}

Table~\ref{tab:backends} summarises the eleven backends benchmarked.
Backend~A terminated before completing 16~qubits due to memory
exhaustion from $\mathcal{O}(4^n)$ Kronecker expansion; all others
completed 3--30 qubits.  All backends use complex64 arithmetic
(8~bytes per amplitude).

\begin{table}[h]
\centering
\caption{Benchmarked Simulation Backends}
\label{tab:backends}
\resizebox{\columnwidth}{!}{%
\begin{tabular}{clllc}
\toprule
\textbf{ID} & \textbf{Framework} & \textbf{Algorithm} &
\textbf{HW} & \textbf{Complexity} \\
\midrule
A & NumPy       & Kronecker (brute-force)  & CPU & $\mathcal{O}(4^n)$ \\
B & pykronecker & Kronecker (lazy)         & CPU & $\mathcal{O}(2^n)$ \\
C & JAX/XLA     & Tensordot $[2]^n$        & CPU & $\mathcal{O}(2^n)$ \\
D & NumPy       & Direct-index             & CPU & $\mathcal{O}(2^n)$ \\
E & NumPy       & Direct-index (NVMe)      & CPU & $\mathcal{O}(2^n)$ \\
F & MLX         & Tensordot $[2]^n$        & GPU & $\mathcal{O}(2^n)$ \\
G & MLX         & Tensordot $[2]^n$        & CPU & $\mathcal{O}(2^n)$ \\
H & MLX         & Flat index $[2^n]$       & GPU & $\mathcal{O}(2^n)$ \\
I & MLX         & Flat index $[2^n]$       & CPU & $\mathcal{O}(2^n)$ \\
J & MLX         & Direct-index             & GPU & $\mathcal{O}(2^n)$ \\
K & MLX         & Direct-index             & CPU & $\mathcal{O}(2^n)$ \\
\bottomrule
\end{tabular}%
}
\end{table}

Backend~E (NVMe) results are excluded from the cliff analysis; its
performance is governed by NVMe bandwidth (${\sim}$3.5~GB/s) rather
than DRAM and is outside the scope of this study.

The \emph{brute-force Kronecker} approach (A) explicitly materialises
the full $2^n \times 2^n$ gate matrix.  The \emph{lazy Kronecker}
approach (B) avoids explicit materialisation.  The \emph{tensordot}
backends (C, F, G) reshape the state vector into $[2,2,\ldots,2]$ and
apply gates as contractions along the target axis, enabling
cache-friendly contiguous block access at the cost of reshape and
transpose overhead.  The \emph{flat-index} backends (H, I) keep the
state vector as $[2^n]$ and use explicit index gathering.

\subsection{Direct-Index Implementation}

The direct-index backends (D, E, J, K) apply gates via bitwise index
manipulation on the flat state vector, computing amplitude pair indices
via XOR and performing in-place scatter-write operations.  For a
single-qubit gate acting on qubit $t$:
\begin{equation}
    b = a \oplus (1 \ll t), \quad \forall\, a : (a \gg t) \& 1 = 0
\end{equation}
For a gate on qubit $t$ the stride between paired elements is $2^t$,
with no reshape or transpose.  The access-pattern implications of this
varying stride are analyzed in Section~\ref{sec:discussion_cliff}.

Backend~D is single-threaded NumPy; backend~J dispatches as parallel
Metal compute shaders.  Backend~E uses the same algorithm on a
memory-mapped NVMe file, extending simulation beyond DRAM capacity.
Our direct-index implementation was developed independently.
The QARN paper~\cite{11333848} describes a conceptually similar
scatter-write approach under the name ARUN (Algorithmic Rearrangement
of Unshuffled Numbers), implemented in Cython with explicit CPU thread
parallelism. However, no independent implementation of ARUN is publicly
available.  Our implementation dispatches the same scatter-write
pattern as parallel Metal compute shaders on GPU (backend~J) and
as sequential NumPy operations on CPU (backend~D).
The implementation is available at
\url{https://github.com/gyanpratipat/qsim-uma}.

\subsection{Experimental Design}

Four benchmark experiments were conducted:
\begin{enumerate}
    \item \textbf{GHZ Statistical (Exp.~1):} All 11 backends, 3--30
    qubits, $N=7$ trials.  Primary algorithm comparison; identifies
    the cliff.

    \item \textbf{GHZ Thermally Isolated (Exp.~2):} Backends C, F, G,
    H, I, J, K; 27--30 qubits; $N=5$ trials; 90~s thermal recovery
    between backends.  Corrects thermal artifacts; provides definitive
    cliff ratios (Table~\ref{tab:cliff}).

    \item \textbf{QFT Single Run (Exp.~3):} All backends, 3--30
    qubits, $N=1$.  Confirms cliff location under a different circuit
    structure.

    \item \textbf{QFT Thermally Isolated (Exp.~4):} Backends C, F, J,
    K; 27--30 qubits; $N=3$ trials; same protocol as Exp.~2.
    Cross-validates cliff circuit-independence
    (Table~\ref{tab:cliff_qft}).
\end{enumerate}

Two further experiments, added to establish causation rather than
description, are described with their results in
Sections~\ref{sec:dtype} and~\ref{sec:crosssim}: a paired
precision probe (Exp.~5) and a cross-simulator comparison against
production state-vector simulators (Exp.~6).

\subsection{Statistical Methodology}

For each (backend, qubit count) pair: one warm-up trial was discarded
to absorb JIT compilation and cache priming; timed trials used
\texttt{time.perf\_counter()}; statistics are mean $\pm 1\sigma$ and
coefficient of variation (CoV $= \sigma/\mu$).  MLX backends ran in
isolated subprocesses (\texttt{multiprocessing} spawn context,
\texttt{maxtasksperchild=1}) to prevent Metal GPU state from persisting
between trials.  Memory is reported as theoretical state vector size
($2^n \times 8$ bytes); XLA and Metal device allocators are invisible
to Python's heap profiler.

\subsection{Thermal Isolation Protocol}

Experiment~1 ran all 11 backends sequentially without thermal recovery.
Post-hoc analysis found cumulative thermal loading inflated JAX CPU
timing by 2.3$\times$ at 28 qubits and 2.8$\times$ at 29 qubits,
producing an artifactual $30q < 29q$ anomaly and a misreported cliff
ratio of 5.28$\times$ instead of the 4.3--4.5$\times$ measured under
thermal isolation.

Experiments~2 and~4 address this with: one backend run completely
before the next begins; 90~s idle sleep between backends;
\texttt{caffeinate -i} active throughout; each MLX backend in a fresh
subprocess.

\subsection{Qubit Range}

Benchmarks span 3 to 30 qubits.  At 32 qubits MLX encounters a uint32
element-count ceiling; at 33+ qubits the state vector exceeds available
DRAM.  Thermally isolated cliff characterization covers 27--30 qubits
only, as thermal artifacts are negligible below 25 qubits.

\section{Results}
\label{sec:results}

\subsection{Arithmetic Intensity and Roofline Position}
\label{sec:roofline_position}

For a single-qubit gate applied to qubit $t$, the direct-index
implementation performs 8~floating-point operations per amplitude
pair across $2^{n-1}$ pairs, transferring 32~bytes per amplitude pair
(16~bytes read~+~16~bytes write, complex64):
\begin{equation}
    \text{AI} = \frac{8~\text{FLOP}}{32~\text{bytes}}
    = 0.25~\text{FLOP/byte}
\end{equation}

Table~\ref{tab:ai} extends this to all standard gate types.
All gates used in this study (H, CNOT, CP, SWAP) have
AI~$\leq 0.38$~FLOP/byte; the general unitary upper bound is
0.875~FLOP/byte.  All values fall deep in the memory-bound regime for
any realistic peak-compute figure on modern hardware.  Compute
throughput is irrelevant to simulation speed; the workload is
structurally memory-bound at all qubit counts.

\begin{table}[h]
\centering
\caption{Arithmetic intensity of standard quantum gates under
direct-index state-vector simulation.
Bytes = DRAM traffic per operation unit (read~+~write).
CP touches only the $|11\rangle$ amplitude;
Phase/P($\theta$) touch only the $|1\rangle$ amplitude.}
\label{tab:ai}
\small
\begin{tabular}{llrrl}
\toprule
\textbf{Gate} & \textbf{Type} & \textbf{FLOPs} &
\textbf{Bytes} & \textbf{AI (F/B)} \\
\midrule
Pauli-X              & 1-qubit &  0 & 32 & 0.000 \\
Pauli-Y              & 1-qubit &  0 & 32 & 0.000 \\
Pauli-Z              & 1-qubit &  0 & 16 & 0.000 \\
CNOT                 & 2-qubit &  0 & 32 & 0.000 \\
CZ                   & 2-qubit &  0 & 16 & 0.000 \\
SWAP                 & 2-qubit &  0 & 32 & 0.000 \\
Hadamard             & 1-qubit &  8 & 32 & 0.250 \\
Phase (T, P($\theta$)) & 1-qubit &  6 & 16 & 0.375 \\
Rotation (Rx, Ry)    & 1-qubit & 12 & 32 & 0.375 \\
Ctrl-Phase           & 2-qubit &  6 & 16 & 0.375 \\
General 2$\times$2   & 1-qubit & 28 & 32 & 0.875 \\
\bottomrule
\end{tabular}
\end{table}

\subsection{The DRAM Bandwidth Cliff}
\label{sec:cliff}

\subsubsection{GHZ Circuit}

\begin{figure*}[t]
  \centering
  \includegraphics[width=\textwidth]{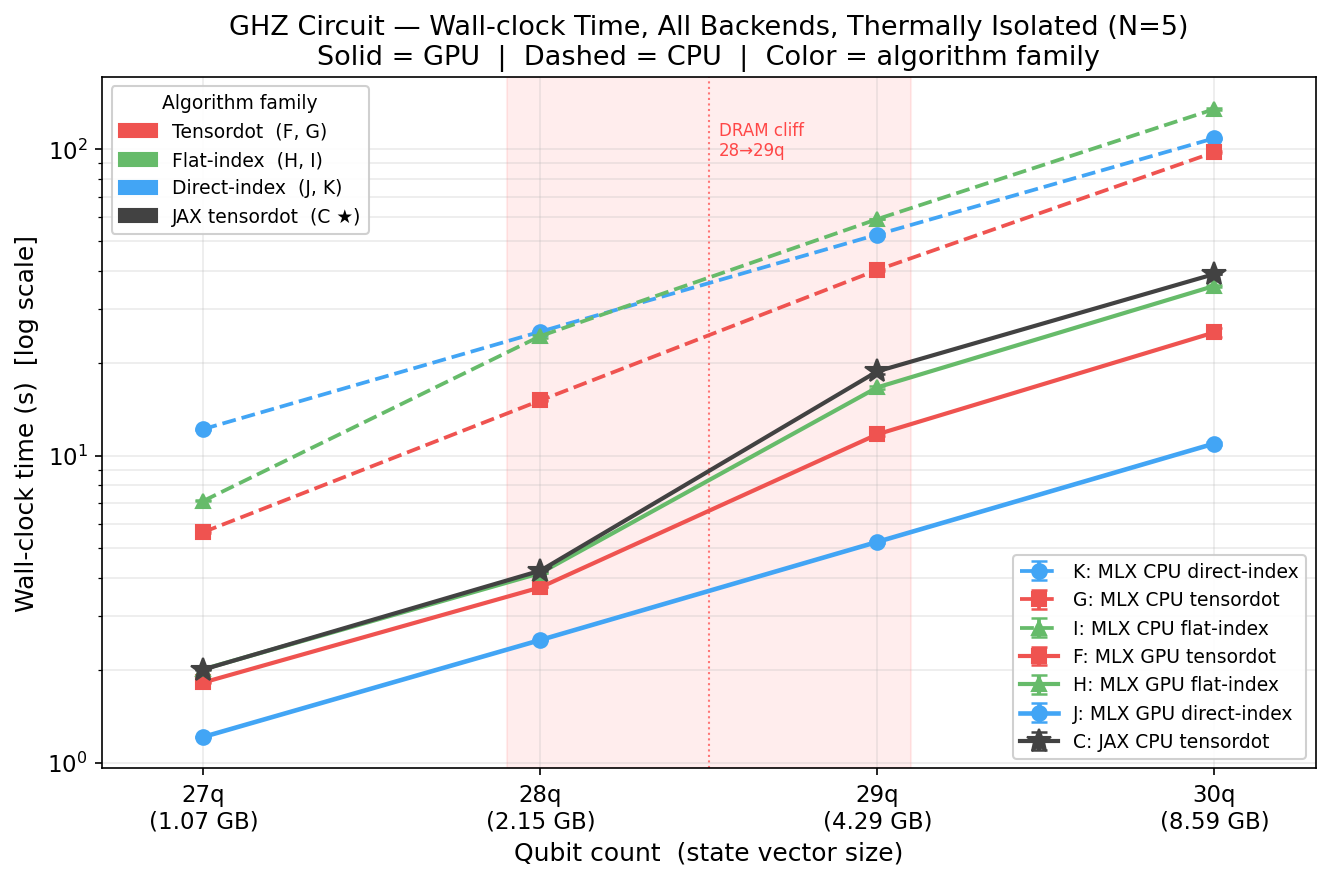}
  \caption{GHZ circuit wall-clock time vs.\ qubit count, all seven backends,
    thermally isolated ($N=5$). Solid lines = GPU; dashed = CPU.
    Color encodes algorithm family: red = tensordot, green = flat-index,
    blue = direct-index, gray = JAX. The shaded band marks the DRAM cliff
    at 28$\rightarrow$29 qubits.}
  \label{fig:ghz_cliff}
\end{figure*}

Figure~\ref{fig:ghz_cliff} shows simulation time versus qubit count for
backends C, F, G, H, I, J, and K across 27--30 qubits (GHZ thermally
isolated, $N=5$).  Backends C and F exhibit a pronounced discontinuity at the
28$\rightarrow$29 qubit transition: runtime increases 4.46$\times$ (C)
and 3.16$\times$ (F), well above the ideal 2$\times$ expected from
$\mathcal{O}(2^n)$ complexity.

\begin{table}[h]
\centering
\caption{Step ratios $t(q)/t(q{-}1)$ at the cliff transition (GHZ
thermally isolated, $N=5$). Ideal doubling $= 2.00\times$.
Bold denotes cliff.}
\label{tab:cliff}
\small
\resizebox{\columnwidth}{!}{%
\begin{tabular}{lccc}
\toprule
\textbf{Backend} & \textbf{28q/27q} & \textbf{29q/28q} &
\textbf{30q/29q} \\
\midrule
C: JAX CPU tensordot & 2.10$\times$ & \textbf{4.46$\times$} & 2.08$\times$ \\
F: MLX GPU tensordot           & 2.04$\times$ & \textbf{3.16$\times$} & 2.15$\times$ \\
G: MLX CPU tensordot           & 2.69$\times$ & 2.66$\times$          & 2.42$\times$ \\
\midrule
H: MLX GPU flat-index          & 2.06$\times$ & \textbf{4.03$\times$} & 2.14$\times$ \\
I: MLX CPU flat-index          & \textbf{3.45$\times$} & 2.41$\times$ & 2.28$\times$ \\
\midrule
J: MLX GPU direct-index        & 2.07$\times$ & 2.09$\times$          & 2.09$\times$ \\
K: MLX CPU direct-index        & 2.08$\times$ & 2.07$\times$          & 2.06$\times$ \\
\bottomrule
\end{tabular}%
}
\end{table}

The cliff marks a step discontinuity in time-per-qubit scaling.  At 28 qubits the state vector
occupies 2.15~GB; although this substantially exceeds the on-chip cache capacity, contiguous
tensor contractions may still benefit from hardware prefetching and access-pattern effects.
At 29 qubits the state vector doubles to 4.29~GB and runtime
more than doubles --- the discontinuity suggests these benefits degrade sharply at this
working-set size.  The same gate operation, on the same algorithm, produces
qualitatively different effective throughput in a single qubit step.

Direct-index backends J and K show no cliff; their step ratios remain
at 2.06--2.09$\times$ throughout.  Flat-index backends H and I show
cliffs at different qubit counts: H at 28$\to$29q (4.03$\times$) and
I at 27$\to$28q (3.45$\times$), producing the mismatched-cliff
artifact discussed in Section~\ref{sec:bandwidth_speedup}.
G (MLX CPU tensordot) shows above-ideal ratios of 2.42--2.69$\times$
throughout the measured window without a single discontinuity,
consistent with G having crossed its cliff before the 27q observation window.  The mechanism is analyzed in
Section~\ref{sec:discussion_cliff}.

\subsubsection{QFT Cross-Validation}

\begin{figure}[t]
  \centering
  \includegraphics[width=\columnwidth]{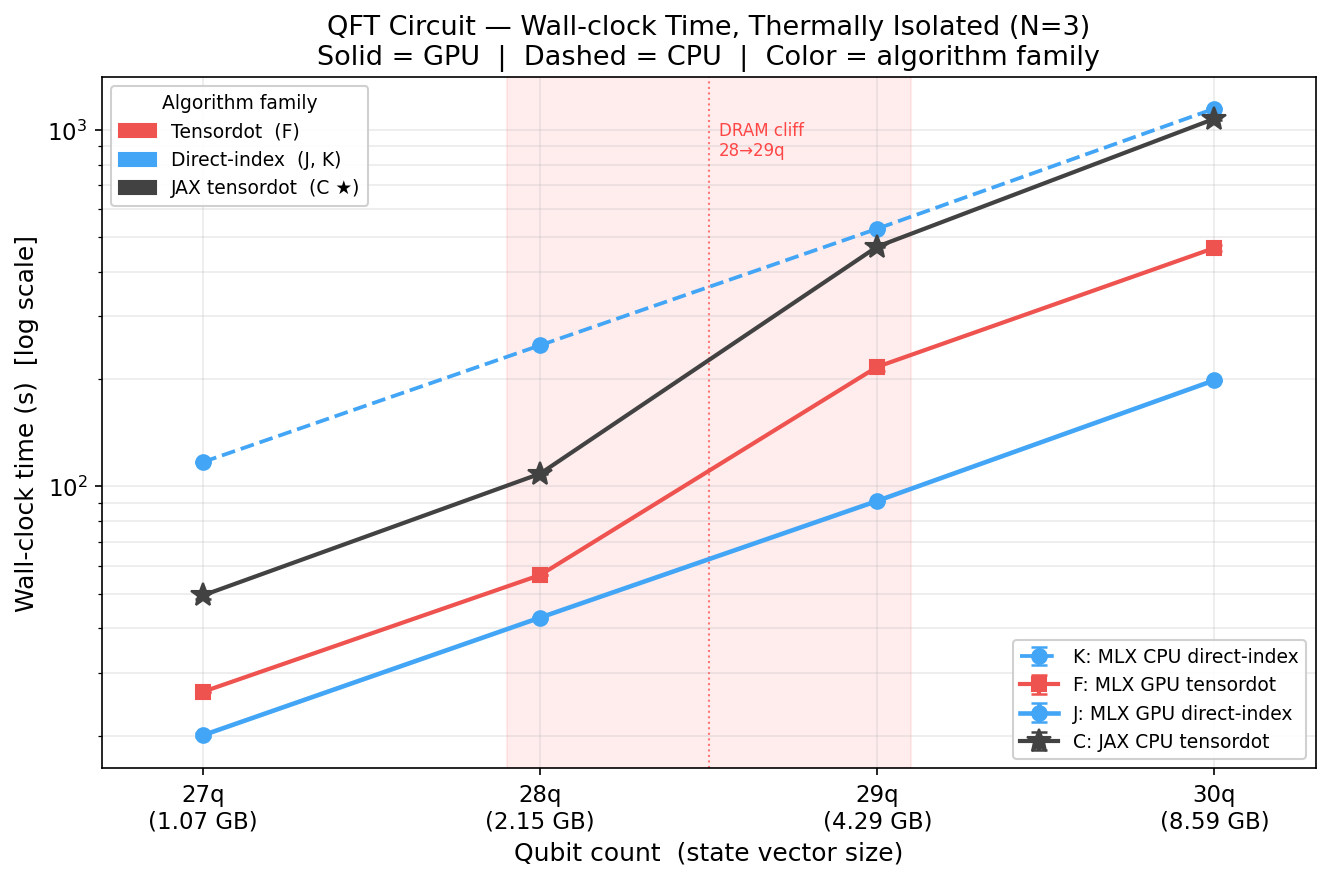}
  \caption{QFT circuit wall-clock time vs.\ qubit count, thermally isolated
    ($N=3$). Same color and line-style conventions as Fig.~\ref{fig:ghz_cliff}.
    The DRAM cliff appears at the same 28$\rightarrow$29 qubit boundary,
    confirming circuit independence.}
  \label{fig:qft_cliff}
\end{figure}

Figure~\ref{fig:qft_cliff} and Table~\ref{tab:cliff_qft} report step
ratios for the QFT thermally isolated experiment ($N=3$, 27--30 qubits).
Backend~C shows a 4.33$\times$ jump at 29q; backend~F shows 3.84$\times$.
Both are consistent with the GHZ cliff ratios.  Backends~J and~K double cleanly.

\begin{table}[h]
\centering
\caption{Step ratios, QFT thermally isolated ($N=3$). Cliff present in
tensordot backends at 29q/28q; absent in direct-index backends. }
\label{tab:cliff_qft}
\small
\resizebox{\columnwidth}{!}{%
\begin{tabular}{lccc}
\toprule
\textbf{Backend} & \textbf{28q/27q} & \textbf{29q/28q} &
\textbf{30q/29q} \\
\midrule
C --- JAX CPU tensordot    & 2.19$\times$ & \textbf{4.33$\times$}         & 2.28$\times$ \\
F --- MLX GPU tensordot    & 2.13$\times$ & \textbf{3.84$\times$} & 2.15$\times$ \\
J --- MLX GPU direct-index & 2.14$\times$ & 2.12$\times$                  & 2.18$\times$ \\
K --- MLX CPU direct-index & 2.13$\times$ & 2.12$\times$                  & 2.17$\times$ \\
\bottomrule
\end{tabular}%
}
\end{table}

The cliff location and the tensordot-versus-direct-index distinction
are circuit-class independent: the same pattern appears under a
fundamentally different gate structure with 16$\times$ higher gate
count (480 gates at 30q versus 30 for GHZ), ruling out GHZ-specific
access patterns as the cause.  The cliff is determined by state vector size and algorithm access pattern,
not by gate count or entanglement depth.
\subsection{Precision as a Causal Probe}
\label{sec:dtype}

Three candidate causes are consistent with the data so far: qubit
count, element count, or working-set size in bytes.  If byte size is
the trigger the discontinuity must move one qubit earlier; if qubit or
element count is, it must stay at 28$\rightarrow$29q.  We re-ran
backend~C (GHZ, thermally isolated, $N=5$) at both precisions and
report the result in Table~\ref{tab:dtype}.

\begin{table}[h]
\centering
\caption{Step ratios indexed by working-set size, backend~C (JAX CPU
tensordot), GHZ, thermally isolated, $N=5$.  The elevated ratio tracks
the 4.29~GB footprint under both precisions, not the qubit count.}
\label{tab:dtype}
\small
\begin{tabular}{lcc}
\toprule
\textbf{Working set} & \textbf{complex64} & \textbf{complex128} \\
\midrule
2.15~GB          & 2.11$\times$ (27$\to$28q)          & 2.10$\times$ (26$\to$27q) \\
\textbf{4.29~GB} & \textbf{4.30$\times$ (28$\to$29q)} & \textbf{5.16$\times$ (27$\to$28q)} \\
8.59~GB          & 2.11$\times$ (29$\to$30q)          & 2.13$\times$ (28$\to$29q) \\
\bottomrule
\end{tabular}
\end{table}

The discontinuity moves exactly one qubit earlier and recurs at the
identical 4.29~GB working set, with clean $\sim$2.1$\times$ doubling on
both sides in both precisions.  A discontinuity remaining at
28$\rightarrow$29q under complex128 would have falsified byte-size
causation and implicated a qubit-count or element-count mechanism;
it did not.  The trigger is therefore the working-set footprint in
bytes, independent of how many qubits or amplitudes produce it.

\subsection{Cross-Simulator Comparison}
\label{sec:crosssim}

If the discontinuity were a property of the hardware, the operating
system, or the memory allocator, every simulator crossing 4.29~GB would
exhibit it.  We tested this against Qiskit Aer~\cite{qiskit2024} and
Qulacs~\cite{qulacs2021}, both multithreaded C++/OpenMP simulators and
far stronger CPU baselines than the single-threaded backends of
Table~\ref{tab:backends}, running the same GHZ circuit on the same
machine.

\begin{table}[h]
\centering
\caption{Step ratio at the 4.29~GB working set across four independent
implementations, GHZ, $N=5$.  Ideal doubling $=2.00\times$.  Only the
kernel that materialises a full transposed buffer departs from it;
susceptibility tracks the memory-access pattern the kernel generates,
not the working-set size alone.}
\label{tab:crosssim}
\small
\resizebox{\columnwidth}{!}{%
\begin{tabular}{llcc}
\toprule
\textbf{Implementation} & \textbf{Access pattern per gate} &
\textbf{c64} & \textbf{c128} \\
\midrule
JAX tensordot (C)   & transpose + full materialised copy & \textbf{4.30$\times$} & \textbf{5.16$\times$} \\
Qiskit Aer          & in-place pair update, gate fusion  & 2.08$\times$ & 2.20$\times$ \\
Qulacs              & in-place, cache-blocked SIMD       & ---          & 1.94$\times$ \\
Direct-index (J, K) & strided gather, DRAM-bound         & 2.09$\times$ / 2.07$\times$ & --- \\
\bottomrule
\end{tabular}%
}
\end{table}

Qulacs allocates well beyond 4.29~GB and doubles cleanly; Aer, which
updates in place and fuses gates, shows at most a mild elevation
(2.20$\times$ in double precision, clean doubling in single).  Only the
tensordot kernel departs from ideal doubling, ruling out a universal
hardware, OS, or allocator threshold.  Byte footprint therefore
determines \emph{where} a susceptible kernel breaks down, while access
pattern determines \emph{whether} it is susceptible at all.

\subsection{STREAM-Style Bandwidth vs.\ Simulation Speedup}
\label{sec:bandwidth_speedup}

\begin{figure}[t]
  \centering
  \includegraphics[width=\columnwidth]{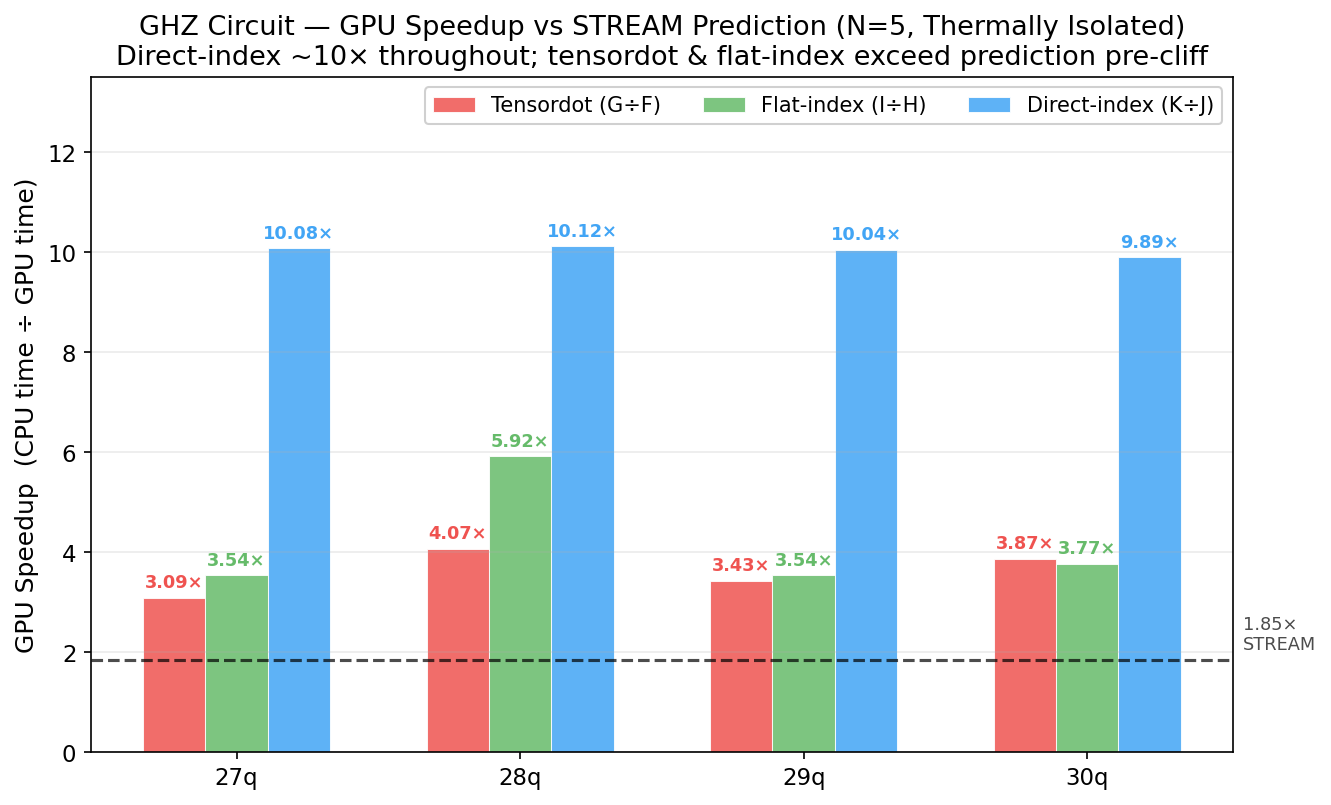}
  \caption{GHZ GPU speedup (CPU time $\div$ GPU time) vs.\ STREAM-style predicted
    1.85$\times$, thermally isolated ($N=5$). Direct-index sustains
    ${\sim}$10$\times$ across all qubit counts; tensordot and flat-index
    exceed the STREAM-style prediction but fall short of direct-index.}
  \label{fig:ghz_speedup}
\end{figure}

\begin{figure}[t]
  \centering
  \includegraphics[width=\columnwidth]{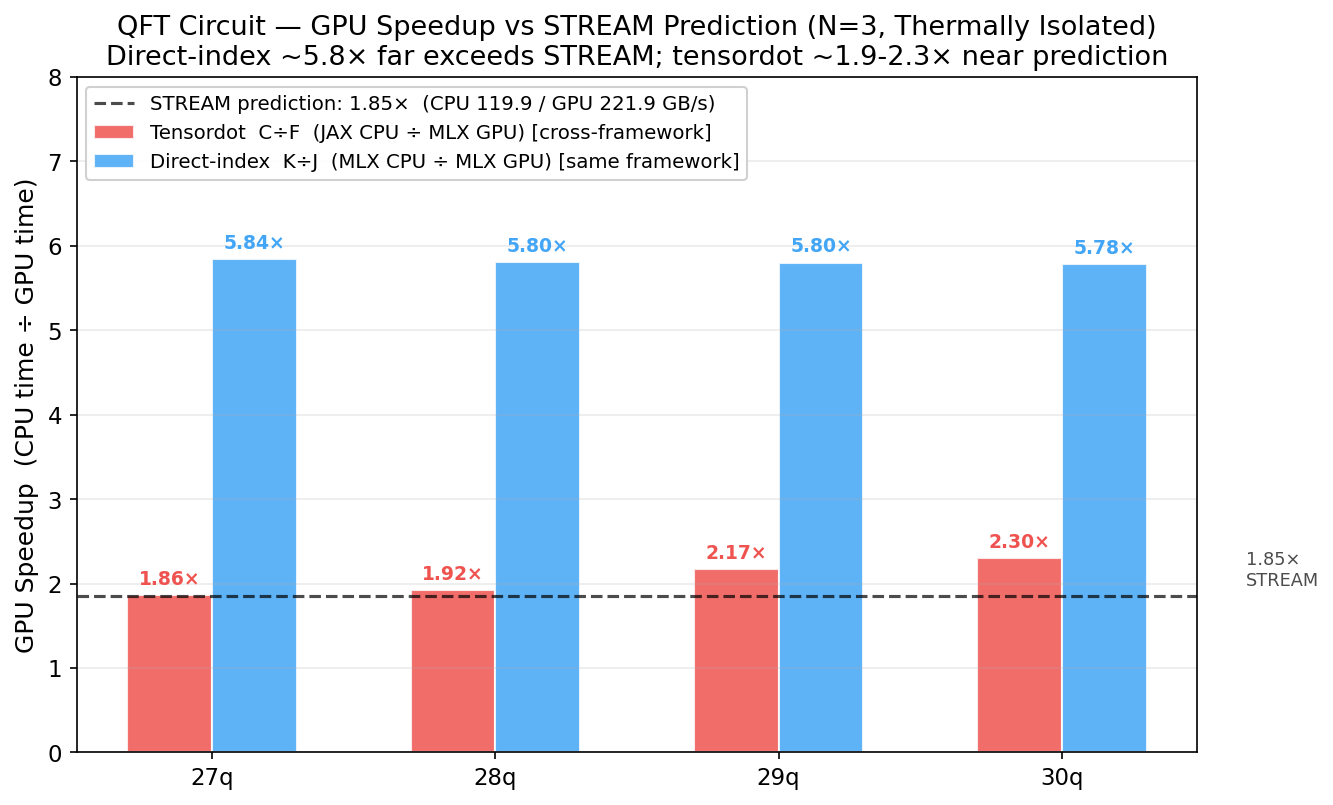}
  \caption{QFT GPU speedup vs.\ STREAM-style predicted 1.85$\times$, thermally
    isolated ($N=3$). Direct-index (K$\div$J, same MLX framework) reaches
    ${\sim}$5.8$\times$. Tensordot (C$\div$F) is a cross-framework
    comparison (JAX CPU vs.\ MLX GPU) and aligns near the STREAM-style
    prediction pre-cliff.}
  \label{fig:qft_speedup}
\end{figure}

Table~\ref{tab:bandwidth} established MLX GPU at 221.9~GB/s and MLX
CPU at 119.9~GB/s, a STREAM-style ratio of 1.85$\times$.  A naive roofline
prediction would therefore expect 1.85$\times$ GPU speedup for all
simulation algorithms.  Table~\ref{tab:speedup} and
Figures~\ref{fig:ghz_speedup}--\ref{fig:qft_speedup} show this is a
significant underestimate for every algorithm tested.

\begin{table}[h]
\centering
\caption{GPU speedup (CPU time / GPU time) across 27--30 qubits (GHZ
thermally isolated, $N=5$).  Values $>1$ mean GPU is faster.  All comparisons use the same MLX
framework; STREAM-style predicted speedup
= 1.85$\times$ (MLX GPU 221.9~GB/s vs.\ MLX CPU 119.9~GB/s) for all
rows.  The flat-index 28q spike reflects mismatched cliff positions:
I crosses its cliff at 27$\to$28q while H crosses at
28$\to$29q, so at 28q the CPU is post-cliff and the GPU is not.}
\label{tab:speedup}
\small
\resizebox{\columnwidth}{!}{%
\begin{tabular}{lccccccc}
\toprule
\textbf{Algorithm} & \textbf{CPU} & \textbf{GPU} &
\textbf{STREAM-style pred.} & \textbf{27q} & \textbf{28q} & \textbf{29q} & \textbf{30q} \\
\midrule
Tensordot    & G & F & 1.85$\times$ & 3.09$\times$ & 4.07$\times$ & 3.43$\times$ & 3.87$\times$ \\
Flat-index   & I & H & 1.85$\times$ & 3.54$\times$ & \textbf{5.92$\times$} & 3.54$\times$ & 3.77$\times$ \\
Direct-index & K & J & 1.85$\times$ & \textbf{10.08$\times$} & \textbf{10.12$\times$} & \textbf{10.04$\times$} & \textbf{9.89$\times$} \\
\bottomrule
\end{tabular}%
}
\end{table}

Three findings emerge from Table~\ref{tab:speedup}:

\textbf{Direct-index (K vs.\ J):} The comparison uses MLX CPU (K)
rather than JAX CPU (C) because framework and algorithm must both be
held constant to isolate the CPU/GPU variable; \texttt{jax-metal}
lacks \texttt{complex64} support (Section~\ref{sec:methodology}), so
no JAX GPU direct-index backend exists.
The STREAM-style measurement predicts 1.85$\times$;
measured speedup is ${\sim}$10$\times$ flat across all qubit counts
(10.08$\times$ at 27q through 9.89$\times$ at 30q).  The consistency
across qubit counts confirms direct-index is DRAM-bound at every
qubit count tested for both CPU and GPU.  The 5.4$\times$ gap beyond
the STREAM-style figure reflects the GPU's massively parallel Metal shader issuing
thousands of concurrent scatter-write operations simultaneously,
versus the CPU's sequential dispatch path---a capability invisible to
sequential STREAM-style measurement.

\textbf{Flat-index (I vs.\ H):} Speedup is 3.54$\times$ at 27q,
spikes to 5.92$\times$ at 28q, then returns to 3.54--3.77$\times$ at
29--30q.  The 28q spike is a mismatched-cliff artifact: I (MLX CPU)
crosses its cliff at 27$\to$28q, while H (MLX GPU) does not
cross until 28$\to$29q.  At 28q the CPU has already crossed its cliff boundary while the GPU has not,
inflating the observed speedup beyond its steady-state value.

\textbf{Tensordot (G vs.\ F):} Speedup ranges from 3.09$\times$ to
4.07$\times$.  G (MLX CPU tensor) shows above-ideal ratios throughout the 27--30q
window, consistent with G having crossed its cliff before the 27q observation window;
F (MLX GPU tensor) crosses its cliff at 28$\to$29q.  The elevated
28q ratio (4.07$\times$) reflects G and F being in different performance regimes
at that qubit count.  At 29--30q, where both have crossed their cliffs,
the speedup settles to 3.43--3.87$\times$.

\section{Discussion}
\label{sec:discussion}

\subsection{STREAM-Style Bandwidth as an Insufficient Predictor}

The collective result is that STREAM-style bandwidth is a necessary but
insufficient predictor for quantum circuit simulation speedup.  Peak streaming
bandwidth does not capture either the parallel request-issuance
capability that governs scatter-write throughput, or the cliff-mismatch effects that arise when CPU and GPU cross their respective
cliff boundaries at different qubit counts.

The findings in Section~\ref{sec:bandwidth_speedup} imply that although the roofline model correctly identifies quantum
simulation as memory-bound, it does not distinguish between access
patterns that are sequential-contiguous (well-characterized by STREAM)
and those that are strided-noncontiguous (characteristic of direct-index
simulation).

The J/K comparison (10.1$\times$ actual vs.\ 1.85$\times$ STREAM-style
prediction) demonstrates that the GPU's parallelism advantage for the
scatter-write pattern is the operative variable, not just the bandwidth
ratio.  This has a practical implication: STREAM benchmarks cannot be
used to predict CPU-to-GPU speedup for quantum simulation workloads
with non-contiguous access patterns.  Hardware selection requires
workload-specific measurements at representative qubit counts.

\subsection{The Cliff as an Access-Pattern Discontinuity}
\label{sec:discussion_cliff}

The 28$\rightarrow$29 qubit discontinuity is not explained by
algorithmic complexity: all backends maintain $\mathcal{O}(2^n)$
working memory and the state vector doubles identically at every qubit
step.  The cliff's magnitude is, however, clearly algorithm-dependent:
tensordot backends exhibit a 4.3--4.5$\times$ discontinuity while
direct-index backends show only ${\sim}$2$\times$ scaling, revealing
that memory access pattern and not computational complexity
determines sensitivity to this discontinuity.

\textbf{Evidence for the cause.}  Sections~\ref{sec:dtype}
and~\ref{sec:crosssim} localize the trigger to working-set bytes and to
the kernel's access pattern rather than to the hardware.  Two further
explanations are excluded: the compiled per-gate kernel is structurally
identical across the transition, and memory pressure remained low with
no compression or swap activity.  Together these point to tensordot's
transpose and materialisation behavior as the source.

\textbf{Why direct-index has no cliff.}  For a gate on qubit $t$,
paired indices are at stride $2^t$.  Across a circuit, gates operate
on all qubits $0$ to $n-1$, producing strides from $2^0$ (contiguous)
to $2^{n-1}$ (2.15~GB at 29 qubits).  Gates on high-index qubits
($t \gtrsim 10$) generate strides far exceeding the 128-byte cache
line, causing cache misses regardless of whether the state vector fits
in cache.  Direct-index kernels thus exhibit highly irregular mixed-stride access patterns
that likely reduce effective cache reuse across all qubit counts, producing
near-uniform scaling behavior with no cliff.

This distinction has a practical consequence not obvious from
complexity analysis: a benchmark at 20--24 qubits cannot predict
29-qubit performance for tensordot backends because the two regimes
have different performance determinants.  Direct-index backends
extrapolate cleanly across the boundary.

\subsection{Circuit Independence}

\begin{figure}[t]
  \centering
  \includegraphics[width=\columnwidth]{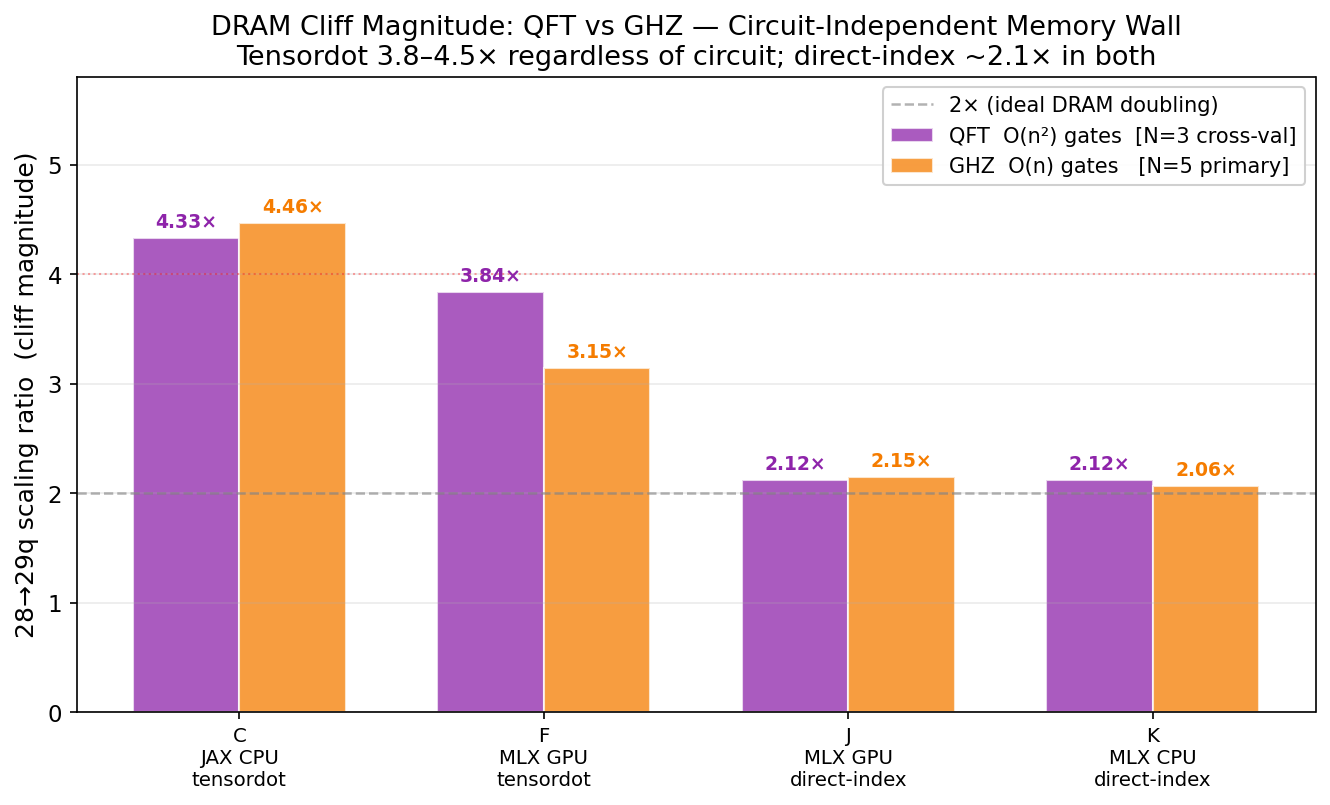}
  \caption{DRAM cliff magnitude (28$\rightarrow$29 qubit step ratio) for QFT
    and GHZ circuits across four backends. Purple = QFT ($N=3$); orange =
    GHZ ($N=5$). Tensordot backends (C, F) cliff at 3.8--4.5$\times$ in
    both circuits; direct-index backends (J, K) remain near 2.1$\times$ in
    both, confirming the cliff is determined by algorithm access pattern,
    not circuit structure.}
  \label{fig:circuit_independence}
\end{figure}

The QFT cross-validation (Table~\ref{tab:cliff_qft},
Fig.~\ref{fig:circuit_independence}) reproduces both the cliff location
and the tensordot-versus-direct-index distinction under
$\mathcal{O}(n)$ and $\mathcal{O}(n^2)$ gate counts.  The result should
extend to any circuit class maintaining a full state vector, including
random circuits, variational ans\"{a}tze, and Grover's algorithm.

\subsection{Thermal Isolation as a Methodological Contribution}

Sequential multi-backend benchmarking without thermal recovery inflated
JAX CPU timing by 2.3$\times$ at 28 qubits and 2.8$\times$ at 29
qubits, producing an artifactual $30q < 29q$ anomaly and misreporting
the cliff ratio as 5.28$\times$ instead of 4.3--4.5$\times$.  Thermal
confounding is particularly acute on fanless or thin-profile systems
under sustained workloads.  The 90-second inter-backend recovery
protocol eliminates the artifact and is reproducible on any platform.

\subsection{Limitations}

MLX GPU backends dispatch Metal compute shaders in parallel, and JAX
applies XLA/AMX multi-threaded compilation.  Production CPU simulators
such as Qiskit~Aer~\cite{qiskit2024} and Qulacs~\cite{qulacs2021}
exploit multi-threaded execution, which would increase CPU effective
bandwidth and narrow the reported CPU-to-GPU ratios.

All simulations use complex64.  Production defaults to complex128,
which doubles memory pressure and shifts the cliff boundary by one
qubit.

Results are from a single hardware platform.  Quantitative bandwidth
values may vary across chip revisions and OS versions.

\subsection{Generalizability Beyond Apple Silicon}

The experimental design is not Apple-specific.  Qualcomm Snapdragon X
Elite~\cite{qualcomm_snapdragon_x_elite}, AMD Ryzen AI
Max~\cite{amd_ryzen_ai_max_uma}, Intel Lunar
Lake~\cite{intel_lunar_lake_uma}, and NVIDIA Grace
Hopper~\cite{nvidia_grace_hopper} share the structural property of a
unified CPU--GPU memory address space.  We make no quantitative claim
about those platforms but provide a paired precision probe and a
cross-simulator comparison under thermal isolation that applies to any
memory-bound workload that may exhibit an access-pattern-dependent
throughput discontinuity.

\section{Conclusion}
\label{sec:conclusion}

We characterized three phenomena in quantum state-vector simulation on
Apple M4~Pro unified memory.  A Roofline analysis places all gates used
in this study at AI $\leq 0.38$~FLOP/byte, establishing structural
memory-boundedness.  A thermally isolated, circuit-independent
throughput discontinuity appears at the 28$\rightarrow$29 qubit
transition, its location fixed by working-set size and its magnitude
4.3--4.5$\times$ in complex64; direct-index
backends show no discontinuity, being uniformly bandwidth-limited
throughout.  And against a STREAM-style prediction of 1.85$\times$,
measured GPU speedups reach 3.1--4.1$\times$ (tensordot),
3.5--5.9$\times$ (flat-index) and 6--10$\times$ (direct-index), so peak
streaming bandwidth does not predict simulation speedup for
non-contiguous access patterns.

The overall insight concerns how a memory-bound, irregular-access
workload behaves on unified-memory hardware.  The contribution is
controlled attribution on a platform where CPU and GPU bandwidth are
identical by measurement rather than by assumption, leaving access
pattern and parallelism as the variables that produce the observed
speedups.  The complex128 shift indicates that the discontinuity is
determined by working-set size, while the comparison against Aer and
Qulacs demonstrates that it is also access-pattern dependent.

\section*{Code Availability}

The simulation backends, benchmark scripts, and thermally isolated
measurement harness described in this paper are available at
\url{https://github.com/gyanpratipat/qsim-uma}.
Reproduction scripts for all experiments and the plotting code for all
figures are included in the repository.

\section*{Acknowledgements}

The author thanks Nishant Kumar Shekhar (Centre for Development of
Advanced Computing, Patna) for reviewing an earlier version of this
manuscript and providing valuable feedback. The author thanks Neh Jigar
Dalal (Arizona State University) for assistance with the PyPI package
release of the companion software. All benchmarks were conducted on an
Apple MacBook Pro (M4~Pro, 48~GB unified memory). The author used large
language model assistants (Claude, Anthropic) for writing feedback and
editorial suggestions during manuscript preparation. The author
acknowledges the open-source communities behind JAX~\cite{jax2018},
MLX~\cite{mlx2023}, and Qiskit~\cite{qiskit2024}.

\bibliographystyle{IEEEtranS}
\bibliography{references}

\end{document}